\documentstyle[11pt,epsfig]{article}
\begin{document}
\newcommand{\be}{\begin{equation}}
\newcommand{\ee}{\end{equation}}
\newcommand{\beq}{\begin{eqnarray}}
\newcommand{\eeq}{\end{eqnarray}}
\newcommand{\ds}{\displaystyle}
\newcommand{\pia}{\mbox{$p_i^{\alpha}$}}
\newcommand{\pjb}{\mbox{$p_j^{\beta}$}}
\newcommand{\la}{\lambda_{\alpha}}
\newcommand{\bla}{\bar{\lambda}_{\alpha}}
\newcommand{\xa}{x^{\alpha}}
\newcommand{\ya}{y^{\alpha}}

\title{Emergence and Growth of Complex Networks in Adaptive Systems}

\author{\hspace{-1.6cm}Sanjay Jain$^1$\footnote{\em jain@cts.iisc.ernet.in}
 and Sandeep Krishna$^2$\footnote{\em sandeep@physics.iisc.ernet.in}\\
\hspace{-1.6cm}$^1${\it Centre for Theoretical Studies, 
Indian Institute of Science, Bangalore 560 012, India} \\
\hspace{-1.6cm}$^2${\it Department of Physics,
Indian Institute of Science, Bangalore 560 012, India }}

\date{}
\maketitle

\vspace{-6cm}     
\rightline{IISc-CTS/10/98}
\vspace{5cm}

\begin{abstract} 
{ We consider the population dynamics of a set of species whose network of
catalytic interactions is described by a directed graph.
The relationship between the attractors of this dynamics and the
underlying graph theoretic structures like cycles and autocatalytic sets
is discussed. It is shown that when the population dynamics is
suitably coupled to a slow dynamics of the graph itself, the network
evolves towards increasing complexity driven by autocatalytic
sets. Some quantitative measures of network complexity are described.
} 
\end{abstract} \vskip 15pt
Complex networks of interacting components are a characteristic feature of
adaptive systems. Examples include prebiotic chemical evolution which 
produced complex organizations of molecules culminating in a living cell,
biological evolution which produces complex ecologies with networks of
interdependent species, and economic and social evolution in which webs
of interacting agents appear spontaneously. 
It is of interest to understand how such webs originate and become more
complex.

In this paper we discuss a model \cite{JK1}\ which describes the
network in terms of a directed graph, and treats it as a dynamical 
variable. The dynamics of the graph is determined by an underlying
population dynamics on a fast time scale in which certain structures,
autocatalytic sets (ACSs) \cite{Kauffman}, play an important role. 
ACSs arise spontaneously in the network and then trigger an increase
in its complexity. 
The main purpose of this paper is to describe the mathematical properties
of ACSs and explain why they cause an increase of complexity of the
network in this model. 

\vskip 2mm
\noindent
{\bf Population dynamics on a fixed artificial chemistry}\\
The system is described by a directed graph with $s$ nodes. 
Associated with each node $i$ is a population $y_i \geq 0$ which evolves in 
time according to
\be
\dot{y}_i = \sum_{j=1}^s c_{ij} y_j - \phi y_i,
\label{ydot}
\ee
where $\phi$ is some function of time and 
$C \equiv (c_{ij})$, $i,j=1, \ldots, s$ is the adjacency matrix of the 
graph, i.e., $c_{ij}=1$ if there is a directed link
from $j$ to $i$ and zero otherwise.
Links from a node to itself are disallowed; $c_{ii}=0$. 

(\ref{ydot}) can be interpreted as a set of rate equations in an 
artificial chemistry similar to the models studied
in \cite{FKP}\cite{FB}\cite{SFM}\cite{SLKP}.
The nodes of the graph correspond to molecular species and a directed link
from node $j$ to node $i$ indicates that species $j$ catalyses the 
production of species $i$.
Let $i$ be produced by a reaction of the type $a+b\stackrel{j}{\rightarrow} i$, 
i.e., from the ligation of reactants $a$ and $b$ catalysed by $j$.
In the approximation of a well stirred chemical 
reactor with a constant dilution flux $\phi$, the rate of growth of $i$ is 
given by $\dot{y}_i=k_f(1+\nu y_j)ab-\phi y_i$, 
where $a$ and $b$ denote reactant populations,
$k_f$ is the rate constant for the spontaneous reaction,
and $\nu$ is the catalytic efficiency \cite{FKP}. 
Assuming that the spontaneous reaction is much slower than the catalysed 
reaction, and that the concentrations of the reactants
are fixed and large, the first term will
be proportional to $y_j$:  
$\dot{y}_i=ky_j-\phi y_i$, $k$ being a constant. 
Making the further idealization that all catalytic strengths are 
equal the rate equations reduce to (\ref{ydot}).

We are interested in the attractors of the relative population dynamics which
follows from (\ref{ydot}), namely
\be
\dot{x}_i = \sum_{j=1}^s c_{ij} x_j -  x_i \sum_{k,j=1}^s c_{kj} x_j,
\label{xdot}
\ee
where $x_i=y_i/ \sum_{j=1}^s y_j$. 
By definition, the relative population vector ${\bf x} \equiv (x_1,\ldots , x_s)$
belongs to the simplex $J \equiv \{ {\bf x} \in
{\bf R}^s | 0 \leq x_i \leq 1, \ \ \sum_{i=1}^s x_i = 1 \} $. 
$J$ is invariant under (\ref{xdot}) since $c_{ij} \geq 0$. 
Let ${\bf X} \equiv (X_1, \ldots , X_s)$ denote an attractive fixed point
of (\ref{xdot}), 
${\bf y}^{\lambda} \equiv (y_1^\lambda, \ldots, y_s^\lambda)$
(or equivalent column vector) a right eigenvector of $C$ with 
eigenvalue $\lambda$, and $\lambda_1$ the eigenvalue of $C$ which has the
largest real part. Then ${\bf x}^{\lambda}={\bf y}^{\lambda}\
/{\sum_{j=1}^s y_j^{\lambda}}$ is a fixed point of (\ref{xdot}).
If $\lambda_1$ is nondegenerate, ${\bf X} = {\bf x}^{\lambda_1}$
is the unique asymptotically stable attractor of (\ref{xdot}) \cite{JK1}.
If $\lambda_1$ is degenerate the attractor configuration is still a 
fixed point; now ${\bf X}$ is a linear superposition of the
${\bf x}^{\lambda_1}$ that depends on the initial condition ${\bf x}(0)$.
The Perron-Frobenius theorem for non-negative matrices 
ensures that $\lambda_1$ is real and $\geq 0$.
Table 1 shows a few simple graphs and their corresponding $\lambda_1$
and attractors. Chains
and trees have $\lambda_1=0$.  Simple cycles
of any size have $\lambda_1=1$ while more complicated structures 
such as the double loop or the eight structure have higher $\lambda_1$.

\vskip 2mm
\noindent
{\bf Autocatalytic sets and spectral properties of directed graphs}\\
An autocatalytic set (ACS) is a subset of the species that 
contains the catalysts for all its members \cite{Kauffman}. In the
present context, one can
define an ACS as a subgraph, each of whose nodes
has at least one incoming link from a node belonging to the
same subgraph. The simplest ACS is a 2-cycle. Every cycle is an
ACS but the converse is not true.  

(i) {\it If a graph has no cycle then $\lambda_1=0$}.
If $C$ is the adjacency matrix of a graph then
$(C^n)_{ij}$ counts the number of distinct paths of length $n$ from node
$j$ to node $i$.
If $\lambda_i$ are the eigenvalues of $C$ then $\lambda_i^n$ are the 
eigenvalues of $C^n$.
If a graph has no cycle then let the length of the longest path between 
any two nodes of the graph be denoted $r$.
Clearly $C^m=0$ for $m>r$. Therefore all eigenvalues of $C^m$ are zero.
Hence, all eigenvalues of $C$ are zero which implies $\lambda_1=0$. 

(ii) {\it If a graph has a cycle then $\lambda_1 \geq 1$}.
If a graph has a cycle then there is some vertex $i$ which has at least 
one path to itself of length $n$,
i.e. $(C^n)_{ii} \geq 1$, for infinitely many values of $n$.
Since the sum of the diagonal entries of a matrix equals the sum of 
the eigenvalues of the matrix, 
$\sum_{i=1}^s (C^n)_{ii}$ is equal to the
sum of the $n^{th}$ powers of the eigenvalues of $C$. 
Thus the sum of the $n^{th}$ powers of the
eigenvalues is at least 1, for infinitely many values of $n$.
Therefore, there exists an eigenvalue with modulus $\geq 1$.
By the Perron-Frobenius theorem, $\lambda_1$ is the eigenvalue with the
largest modulus, hence $\lambda_1 \geq 1$ \cite{BH}.

An ACS is not in general an irreducible
subgraph, because there need not be a path from every node of an
ACS to another. Thus the Perron-Frobenius theorem for irreducible
non-negative matrices \cite{Seneta} does not
apply to ACSs. Nevertheless, ACSs share some important properties
with cycles (which are irreducible): 

(iii) {\it An ACS must contain a cycle}.
Let $C$ be the adjacency matrix of a graph which is itself an ACS. 
Then by definition,
every row of $C$ has at least one non-zero entry.
Construct $C'$ by removing all non-zero entries in each row of $C$
except one which can be chosen arbitrarily.
Thus $C'$ has exactly one non-zero entry in each row.
Clearly the column vector ${\bf x}=(1,1, \ldots, 1)$ is a right 
eigenvector of $C'$ with eigenvalue 1.
This implies that the graph for which $C'$ is the adjacency matrix contains
a cycle. Since the construction of $C'$ from $C$ involved only removal of 
some links, it follows that the original graph must also contain a cycle.

(iv) {\it If a graph has no ACS then $\lambda_1=0$}. This follows from 
(i) and (iii).

(v) {\it If a graph has an ACS then $\lambda_1 \geq 1$}. This follows from 
(ii) and (iii).

The reason why an ACS is a useful concept in the present context is 
the following property, not true of cycles in general:

(vi) {\it If a graph has $\lambda_1 \geq 1$, then the subgraph
corresponding to the set of nodes $i$ for which $x_i^{\lambda_1} > 0$
is an ACS}. 
Renumber the nodes of the graph so that $x_i^{\lambda_1}>0$ only for 
$i=1, \ldots, k$. Let $C$ be the adjacency matrix of this graph.
Since ${\bf x}^{\lambda_1}$ is an eigenvector of the matrix $C$ we have 
$\sum_{j=1}^s c_{ij}x^{\lambda_1}_j=\lambda_1 x^{\lambda_1}_i
\Rightarrow\sum_{j=1}^k c_{ij}x^{\lambda_1}_j=\lambda_1 x^{\lambda_1}_i$.
Since $x^{\lambda_1}_i>0$ only for $i=1, \ldots, k$ it follows that for each 
$i \in \{1,\ldots,k\}$ there exists a $j$ such that $c_{ij}>0$.
Hence the $k \times k$ submatrix $C' \equiv (c_{ij})$, $i,j=1,\ldots,k$ 
has at least one non-zero entry
in each row. Thus each node of the subgraph corresponding to this submatrix 
has an incoming link from one of the other nodes in the subgraph. Hence the 
subgraph is an ACS.
We call this subgraph the `dominant ACS' of the graph.

(vii) Consider a graph with no cycles and let there be a chain of $r$ links in 
this graph whose successive nodes are labelled $i=1,2, \ldots, r+1$. 
The node 1 (to which there is no incoming link) has
a constant population $y_1$ since the r.h.s of (\ref{ydot}) vanishes for
$i=1$ when $\phi=0$. For the node 2, we get $\dot{y_2}=y_1$, hence
$y_2(t)=y_2(0)+y_1t \sim t$ for large $t$. Similarly it can be
seen that $y_k$ grows as $t^{k-1}$. In general, it is clear that for a graph with
no cycles, $y_i \sim t^r$ for large $t$ when $\phi = 0$, where
$r$ is the length of the longest path terminating at node $i$. 
Since the dynamics (\ref{xdot}) does not depend upon the choice of $\phi$,
this proves that {\it for a graph with no cycles $X_i = 0$ for all $i$ except 
the nodes at which the longest paths in the graph terminate}. 
Similarly if a 2-cycle feeds into another 2-cycle as shown in Table 1, 
$X_i=0$ for the nodes in the first 2-cycle. These are examples of the
more general situation where $X_i=0$ for nodes of a subgraph which
has another subgraph `downstream' from it with an equal or larger 
$\lambda_1$. 

\vskip 2mm
\noindent
{\bf Evolution of the network} \\
On a short timescale the graph remains fixed and the $x_i$ evolve  
according to (\ref{xdot}). On a longer timescale 
at discretely spaced intervals (labelled by $n=1,2,\ldots$)
the graph itself changes by the elimination of an existing species 
and the creation of a new one.
At the initial time ($n=0$) the graph is random:
$c_{ij}=1$ (for $i \neq j$)
with probability $p$ ($p$ is called the `catalytic probability')
and zero with probability $1-p$. 
$m \equiv p(s-1)$ is referred to as the average connectivity. 
The graph at $n+1$ is obtained from that at $n$ by the
following procedure: Determine the `set of least fit nodes' at $n$,
namely, those that have the smallest $X_i$ for the graph at $n$. Pick 
a node at random from this set, and reassign links between it and
all other nodes randomly with the same probability $p$.
The idea is that for a fixed graph the relative population stabilizes
at ${\bf X}$. One of the least fit species ($X_i$ is taken to be a
measure of fitness) mutates or is eliminated. 
This corresponds to an `extremal dynamics' as in the model of 
Bak and Sneppen \cite{BS}.
The graph update
corresponds to the appearance of the mutant or replacement, which
has random connections (but with the same average connectivity $m$ as
in the initial graph) with other species.

Define $s_1(n)$ as the number of species $i$ for which 
$X_i \neq 0$ at the $n^{th}$ time step. 
For a graph with $\lambda_1 \geq 1$, $s_1$ is the number of nodes in 
the dominant ACS.
Whenever $s_1<s$ it can be shown that the set of least fit nodes, 
which is the set of nodes with $X_i=0$,
is unique and independent of the initial condition even if $\lambda_1$
is degenerate. 
When $s_1=s$, $\lambda_1$ has turned out to be nondegenerate in
the runs displayed. 
Hence there is no ambiguity in the update procedure arising from
initial conditions on ${\bf x}$.

Figures 1 and 2 show how $\lambda_1$ and $s_1$ evolve for a run
with $s=100$ and $m=0.25$.
For $n < n_1 = 1515$, $\lambda_1=0$ and 
there is no ACS in the graph. 
For $n<n_1$, $s_1$ is below 10 most of the time. 
At $n=n_1$ the graph update forms a 3-cycle 
which becomes the dominant ACS and 
$\lambda_1$ jumps from zero to one. 
The nodes outside the dominant ACS by definition
have $x_i^{\lambda_1} = 0$ and 
constitute the set of least fit nodes. Therefore,
as long as the dominant ACS at step $n$ (let us denote this subgraph
as A($n$)) does not include the
whole graph, i.e., $s_1<s$, the mutating node will be outside it.
At step $n+1$ the mutant species can either (a) 
get linked to A($n$), or 
(b) form another ACS with other nodes which were not part of 
A($n$), or (c) be a singleton or part of a non-ACS structure.
In all cases $\lambda_1(n+1)$ cannot be less than $\lambda_1(n)$.
(For case (a) this depends upon the fact that the mutating node,
being outside A($n$), cannot destroy any of its links.) Thus,
{\it whenever $s_1 < s$, $\lambda_1$ is a non-decreasing function
of $n$}. It follows that once an ACS is formed by chance, the
autocatalytic property of the graph will be preserved 
until the dominant ACS engulfs the whole graph.

After the appearance of the ACS at $n=n_1$, 
whenever a node gets an incoming link from the dominant ACS
it becomes part of the dominant ACS. Most of the time this increases $s_1$ 
and the dominant ACS grows until it spans the
entire graph at $n=n_2=3010$ when $s_1=s$ for the first time.
For $n \in [n_1,n_2]$, $s_1$ locally averaged in time grows exponentially.

In the given run a 2-cycle 
disconnected from the existing dominant ACS
is created $n=1807$. Both the cycles coexist and
grow together until $n=2145$ when the graph update creates a chain
from the 3-cycle to the 2-cycle.
Thus a situation occured in which a subgraph with $\lambda_1=1$, comprising
the 3-cycle and the nodes being fed by the 3-cycle,
was now feeding into the 2-cycle `downstream' which also had $\lambda_1=1$.
Hence all the nodes of the first subgraph 
entered the set of least fit nodes (see remarks under (vii)) 
and $s_1$ decreased from 15 to 10. 
Later at $n=2240$ the chain joining the cycles was broken by the graph
update and both cycles became part of the dominant ACS.
At $n=2607$, $s_1$ decreases from 58 to 31. This time also a chain formed 
between the cycles with the 2-cycle now feeding into the 3-cycle.
$s_1$ rises back up to $54$ at $n=2611$ when this chain too was broken
by the graph update. There is one more fall in $s_1$ at $n=2780$. The graph
update results in the 3-cycle being converted to a double loop which has
$\lambda_1\approx 1.17$ which overshadows the ACS whose core is the 2-cycle.
 
Even in regions where $s_1$ is increasing the ACS structure could be changing
quite a lot.
Many complicated processes are possible, 
such as formation of new disconnected ACSs or the 
reinforcement of an old overshadowed ACS before it was completely broken up.
Thus the actual growth
of the ACS is usually a very complicated process with the dominant
ACS often undergoing drastic changes in structure caused by purely 
chance events. We wish to emphasize, however, that notwithstanding the
historical particularities of a given run, 
every history respects the rule that 
$\lambda_1$ is a nondecreasing function of time (unless $s_1 = s$)
in this simple model. (In a `non-extremal' dynamics, one can expect
a stochastic version of this monotonicity, as long as selection
is still sufficiently strong.) Further, 
the {\it ensemble} of runs 
with the same parameter values $m$, $s$ has identifiable 
characteristics like the average time of appearance of an ACS
$\tau_a \equiv \langle n_1 \rangle \sim s/m^2 \sim 1/(p^2 s)$, and
the exponential growth time scale $\tau_g \sim s/m \ 1/p$ 
(for sufficiently small $m$ and large $s$) \cite{JK1}. 

At $n=n_2$ the whole graph becomes an ACS and for the
first time the graph update will alter a node from the dominant ACS.  
Then $\lambda_1$ can decrease. It eventually settles in a statistical
steady state with large fluctuations in the run shown. For very low $m$,
$\lambda_1$ can even become zero in a graph update. 

In addition to $\lambda_1$, $s_1$ and the total number of links $l$, another
measure of the complexity of the graph is
`interdependency', denoted $\bar{d}$,
which we define as follows:
$\bar{d} \equiv (1/s)\sum_{i=1}^s d_i$, where $d_i$ is the
`dependency' of the $i^{th}$ node.
$d_i$ is the total number of links
in all paths that terminate at node $i$, each link counted only once.
Since $d_i$ counts how many
links ultimately `feed into' the node $i$, it is a  measure of
how `dependent' species $i$ is on other species. Thus $\bar{d}$ is
a measure of how interdependent the species in the graph are.  From 
Table 1 it is evident 
that graphs with the same number of links 
can have different $\bar{d}$.

Figure 3 shows $\bar{d}$ versus $n$ for the same run 
(the lower curve in the figure corresponds to a `random run' in which the 
node to be eliminated is picked randomly from the entire set of $s$ nodes).
The qualitative features are similar to $l$ which has been discussed
in \cite{JK1}. A quantitative difference is that 
from $n=0$ to the steady state, $\bar{d}$ increases 
by a factor of approximately 50 which is much more than 
the increase in $l$ (a factor of about 5, see \cite{JK1}). 
The evolution of the system does not seem to lead to maximally
dense graphs, $m^* \sim s$, but only $m^* \sim O(1)$ 
($m^*$ is the steady state connectivity). Nevertheless the 
system achieves a high interdependency.  
A steady state $\bar{d}$ value of about 25 
means that each species is being fed with an appreciable fraction of
the total number of links, suggestive of long range correlations
having developed in the system.  The fluctuations of $\bar{d}$ in the 
steady state are also more pronounced than for $l$. 
At each graph update the change in the number
of links is $O(1)$ for fixed $m$ but the addition or removal of a few links
causes large changes in $\bar{d}$, reflecting its nonlocal character.

\noindent {\bf Acknowledgement}:
S. J. acknowledges the affiliation and
support of the Jawaharlal Nehru Centre for Advanced Scientific Research,
Bangalore, as well as Associate membership 
and hospitality of the Abdus Salam International
Centre for Theoretical Physics, Trieste.

\begin{figure}[h]
\epsfxsize=8cm
\centerline{\epsfbox{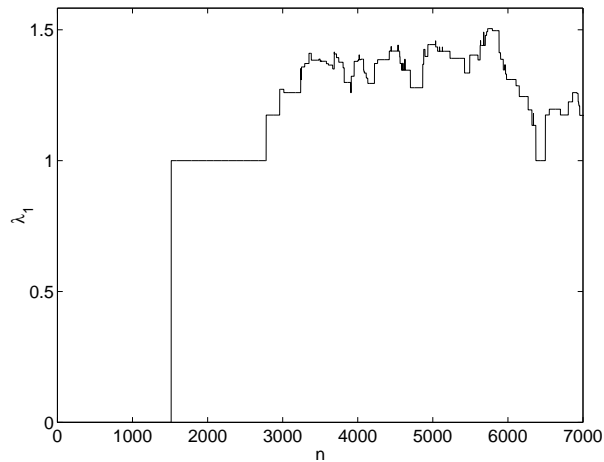}}
\caption{$\lambda_1$ versus $n$ for $s=100, m=0.25$.} 
\end{figure}

\begin{figure}
\epsfxsize=8cm
\centerline{\epsfbox{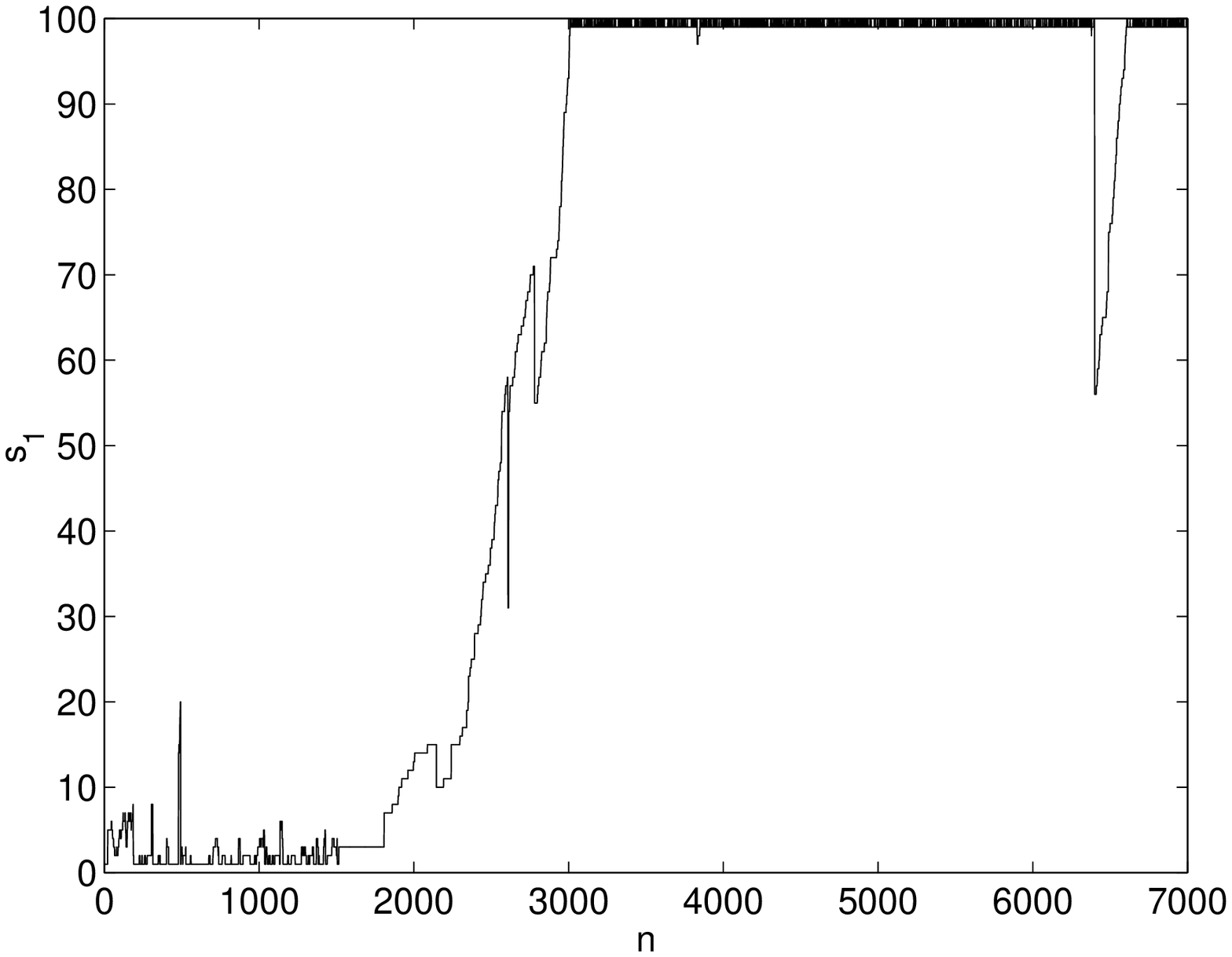}}
\caption{$s_1$ versus $n$ for $s=100, m=0.25$.} 
\end{figure}

\begin{figure}
\epsfxsize=8cm
\centerline{\epsfbox{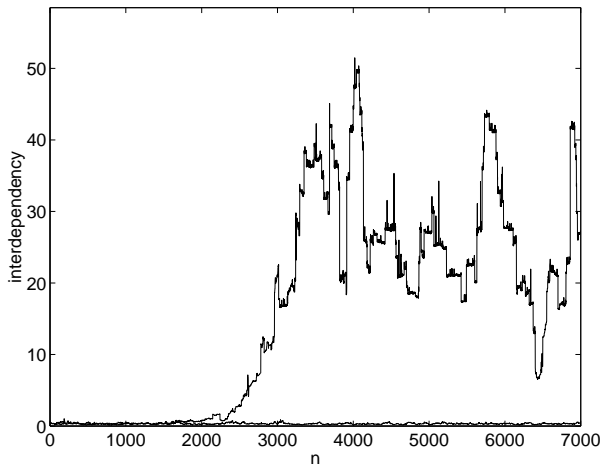}}
\caption{Interdependency versus $n$ for $s=100, m=0.25$.
The lower curve is for a random run with $s=100, m=0.25$.}
\end{figure}

\pagebreak
\begin{table}
\vspace{-2cm}
\caption{Adjacency matrices, largest eigenvalues
, the corresponding eigenvectors and interdependency of some simple graphs}
\begin{tabular}{|c|c|c|c|c|}
\hline
Graph & C & $\lambda_1$ & ${\bf x}^{\lambda_1}$ & $\bar{d}=(d_1+d_2+ \ldots +d_s)/s$ \\
\hline
\begin{picture}(100,40)
\put(35,20){chain}
\put(5,5){\circle*{5}}
\put(2.5,-8){1}
\put(5,5){\vector(1,0){27.5}}
\put(35,5){\circle*{5}}
\put(32.5,-8){2}
\put(35,5){\vector(1,0){27.5}}
\put(65,5){\circle*{5}}
\put(62.5,-8){3}
\put(65,5){\vector(1,0){27.5}}
\put(95,5){\circle*{5}}
\put(92.5,-8){4}
\end{picture} 
& 
$\left( \begin{array}{cccc}
0 & 0 & 0 & 0\\
1 & 0 & 0 & 0\\
0 & 1 & 0 & 0\\
0 & 0 & 1 & 0\\
\end{array} \right)$ &
0 &
$\left( \begin{array}{c}
0 \\
0 \\
0 \\
1 \\
\end{array} \right)$ &
 1.5=(0+1+2+3)/4\\
\hline
\begin{picture}(100,40)
\put(30,30){tree}
\put(80,-15){\vector(-3,1){57.5}}
\put(80,5){\vector(-1,0){57.5}}
\put(80,25){\vector(-3,-1){57.5}}
\put(20,5){\circle*{5}}
\put(10,2){1}
\put(80,25){\circle*{5}}
\put(85,22){2}
\put(80,-15){\circle*{5}}
\put(85,-18){4}
\put(80,5){\circle*{5}}
\put(85,2){3}
\end{picture}
&
$\left( \begin{array}{cccc}
0 & 1 & 1 & 1\\
0 & 0 & 0 & 0\\
0 & 0 & 0 & 0\\
0 & 0 & 0 & 0\\
\end{array} \right)$ &
0 & 
$\left( \begin{array}{c}
1 \\
0 \\
0 \\
0 \\
\end{array} \right)$ &
0.75=(3+0+0+0)/4\\
\hline
\begin{picture}(100,40)
\put(33,30){2-cycle}
\put(20,20){\circle*{5}}
\put(10,17){1}
\put(20,20){\line(0,-1){15}}
\put(30,5){\oval(20,20)[bl]}
\put(30,-5){\vector(1,0){47.5}}
\put(80,-5){\circle*{5}}
\put(85,-8){2}
\put(80,-5){\line(0,1){15}}
\put(70,10){\oval(20,20)[tr]}
\put(70,20){\vector(-1,0){47.5}}
\end{picture}
&
$\left( \begin{array}{cc}
0 & 1\\
1 & 0\\
\end{array} \right)$ &
1 & 
$\left( \begin{array}{c}
1/2 \\
1/2 \\
\end{array} \right)$ &
2=(2+2)/2\\
\hline
\begin{picture}(100,40)
\put(5,30){2-cycle feeding into}
\put(60,20){a node}
\put(20,20){\circle*{5}}
\put(10,17){1}
\put(20,20){\line(0,-1){15}}
\put(30,5){\oval(20,20)[bl]}
\put(30,-5){\vector(1,0){17.5}}
\put(50,-5){\circle*{5}}
\put(47.5,-15){2}
\put(50,-5){\line(0,1){15}}
\put(40,10){\oval(20,20)[tr]}
\put(40,20){\vector(-1,0){17.5}}
\put(52.5,-5){\vector(1,0){25}}
\put(80,-5){\circle*{5}}
\put(85,-8){3}
\end{picture}
&
$\left( \begin{array}{ccc}
0 & 1 & 0\\
1 & 0 & 0\\
0 & 1 & 0\\
\end{array} \right)$ &
1 & 
$\left( \begin{array}{c}
1/3 \\
1/3 \\
1/3 \\
\end{array} \right)$ &
2.33=(2+2+3)/3\\
\hline
\begin{picture}(100,50)
\put(5,40){2-cycle feeding into}
\put(30,30){a 2-cycle}
\put(20,20){\circle*{5}}
\put(10,17){1}
\put(20,20){\line(0,-1){15}}
\put(30,5){\oval(20,20)[bl]}
\put(30,-5){\vector(1,0){7.5}}
\put(40,-5){\circle*{5}}
\put(37.5,-18){2}
\put(40,-5){\line(0,1){15}}
\put(30,10){\oval(20,20)[tr]}
\put(30,20){\vector(-1,0){7.5}}

\put(42.5,-5){\vector(1,0){15}}

\put(80,20){\circle*{5}}
\put(85,17){4}
\put(80,20){\line(0,-1){15}}
\put(70,5){\oval(20,20)[br]}
\put(70,-5){\vector(-1,0){7.5}}
\put(60,-5){\circle*{5}}
\put(57.5,-18){3}
\put(60,-5){\line(0,1){15}}
\put(70,10){\oval(20,20)[tl]}
\put(70,20){\vector(1,0){7.5}}
\end{picture}
&
$\left( \begin{array}{cccc}
0 & 1 & 0 & 0\\
1 & 0 & 0 & 0\\
0 & 1 & 0 & 1\\
0 & 0 & 1 & 0\\

\end{array} \right)$ &
1 & 
$\left( \begin{array}{c}
0 \\
0 \\
1/2 \\
1/2 \\
\end{array} \right)$ &
3.5=(2+2+5+5)/4\\
\hline
\begin{picture}(100,40)
\put(35,30){3-cycle}
\put(20,20){\circle*{5}}
\put(10,17){1}
\put(20,20){\line(0,-1){15}}
\put(30,5){\oval(20,20)[bl]}
\put(30,-5){\vector(1,0){17.5}}
\put(50,-5){\circle*{5}}
\put(47.5,-15){2}
\put(50,-5){\line(1,0){20}}
\put(70,5){\oval(20,20)[br]}
\put(80,5){\vector(0,1){12.5}}
\put(80,20){\circle*{5}}
\put(85,18){3}
\put(80,20){\vector(-1,0){57.5}}
\end{picture}
&
$\left( \begin{array}{ccc}
0 & 0 & 1\\
1 & 0 & 0\\
0 & 1 & 0\\
\end{array} \right)$ &
1 & 
$\left( \begin{array}{c}
1/3 \\
1/3 \\
1/3 \\
\end{array} \right)$ &
3=(3+3+3)/3\\
\hline
\begin{picture}(100,40)
\put(38,30){eight}
\put(20,25){\circle*{5}}
\put(10,22){1}
\put(20,25){\line(0,-1){20}}
\put(30,5){\oval(20,20)[bl]}
\put(30,-5.1){\vector(1,0){17.5}}
\put(50,-5){\circle*{5}}
\put(47.5,-15){2}
\put(50,-5.1){\line(1,0){20}}
\put(70,5){\oval(20,20)[br]}
\put(80,5){\vector(0,1){17.5}}
\put(80,25){\circle*{5}}
\put(85,22){3}
\put(80,25){\vector(-1,-1){27.5}}
\put(50,-5){\vector(-1,1){27.5}}
\end{picture}
&
$\left( \begin{array}{ccc}
0 & 1 & 0\\
1 & 0 & 1\\
0 & 1 & 0\\
\end{array} \right)$ &
$\sqrt{2}$ & 
$\left( \begin{array}{c}
0.293 \\
0.414 \\
0.293 \\
\end{array} \right)$ &
4=(4+4+4)/3\\
\hline
\begin{picture}(100,40)
\put(25,30){double loop}
\put(20,20){\circle*{5}}
\put(10,17){1}
\put(80,20){\circle*{5}}
\put(85,17){4}
\put(20,-10){\circle*{5}}
\put(10,-13){2}
\put(80,-10){\circle*{5}}
\put(85,-13){3}
\put(20,20){\vector(0,-1){27.5}}
\put(20,-10){\vector(1,0){57.5}}
\put(80,-10){\vector(0,1){27.5}}
\put(80,20){\vector(-1,0){57.5}}
\put(80,20){\vector(-2,-1){57.5}}
\end{picture}
&
$\left( \begin{array}{cccc}
0 & 0 & 0 & 1\\
1 & 0 & 0 & 1\\
0 & 1 & 0 & 0\\
0 & 0 & 1 & 0\\
\end{array} \right)$ &
1.22 & 
$\left( \begin{array}{c}
0.18 \\
0.33 \\
0.27 \\
0.22 \\
\end{array} \right)$ &
5=(5+5+5+5)/4\\
\hline
\end{tabular}
\end{table}

\end{document}